\documentclass[prl , twocolumn, superscriptaddress, showpacs,amsmath,amssymb]{revtex4}

\usepackage{float}
\usepackage{graphicx}
\usepackage{subfigure}
\usepackage{verbatim}

\begin{document}

\title{Observation of the dynamic Jahn-Teller effect in the excited states of nitrogen-vacancy centers in diamond}
\author{Kai-Mei C. Fu}
\email{kai-mei.fu@hp.com}
\author{Charles Santori}
\author{Paul E. Barclay}
\affiliation{Information and Quantum Systems Lab, HP Labs, 1501 Page Mill Road, Palo Alto, CA
94304, USA}
\author{Lachlan J. Rogers}
\author{Neil B. Manson}
\affiliation{Laser Physics Center, RSPE, Australian National University, Canberra, ACT 0200, Australia}
\author{Raymond G. Beausoleil}
\affiliation{Information and Quantum Systems Lab, HP Labs, 1501 Page Mill Road, Palo Alto, CA
94304, USA}

\begin{abstract}
The optical transition linewidth and emission polarization of single nitrogen-vacancy (NV) centers are measured from 5~K to room temperature.  Inter-excited state population relaxation is shown to broaden the zero-phonon line and both the relaxation and linewidth are found to follow a $T^5$ dependence for $T$ $<$ 100~K.  This dependence indicates that the dynamic Jahn-Teller effect is the dominant dephasing mechanism for the NV optical transitions at low temperatures.
\end{abstract}

\maketitle

The negatively-charged nitrogen-vacancy (NV) center in diamond has attracted much scientific interest due to its unique spin and optical properties. Long coherence lifetimes of the ground electron spin state at room temperature, up to milliseconds in recent reports ~\cite{ref:Balasubramanian2009usc}, have enabled fundamental studies of coherent electron-electron~\cite{ref:Gaebel2006rtc} and electron-nuclear coupling~\cite{ref:GurudevDutt2007qua,ref:Neuman2008mea,ref:Hanson2008cds} of a small number of spins in a solid. Spin-selective optical transitions enable readout of the electron spin at room temperature which is critical for both high sensitivity magnetometry~\cite{ref:Taylor2008hig} and quantum information processing.  Moreover, at cryogenic temperatures, it is proposed that these spin-dependent optical transitions~\cite{ref:Santori2006cpt, ref:Tamarat2008sfs} could provide an interface between spins and photons as needed in schemes for scalable quantum computation~\cite{ref:Benjamin2006bgs,ref:Clark2007qcb} and quantum communication~\cite{ref:Ladd2006hqr,ref:Childress2005ftq}.  The success of such schemes depends on the coherence properties of the emitted photons which in turn are determined by the coherence properties of the NV excited states.

Dephasing of the excited state of an optical transition manifests itself in an energy broadening of the transition.  For solid-state defects, the broadening can be observed in the zero-phonon line (ZPL), the optical transition in which no net phonon is emitted or absorbed. Typically the ZPL width exhibits a $T^7$ dependence on temperature $T$, and the dephasing mechanism is attributed to electron-phonon scattering mediated by a quadratic electron-phonon interaction\cite{ref:Davies1974vsd, ref:Maradudin1966pdd}. In this work we first measured the polarization of the emitted ZPL photons as a function of $T$. The NV excited state consists of an orbital doublet and a change in the emitted polarization indicates population transfer between the two orbital states. We found that the emission polarization is strongly temperature dependent and that the population relaxation rate between the orbital states varies as $T^5$.  Next we measured the temperature dependence of the ZPL width.  A $T^5$ dependence was again found, indicating that for $T<100$~K the ZPL broadening is mainly determined by excited state population transfer.  Most significantly, the $T^5$ dependence observed is evidence of the dynamic Jahn-Teller (DJT) effect in the NV excited states, as explained below.
\begin{figure}
\centering
{\includegraphics[width = 3.3in,
keepaspectratio]{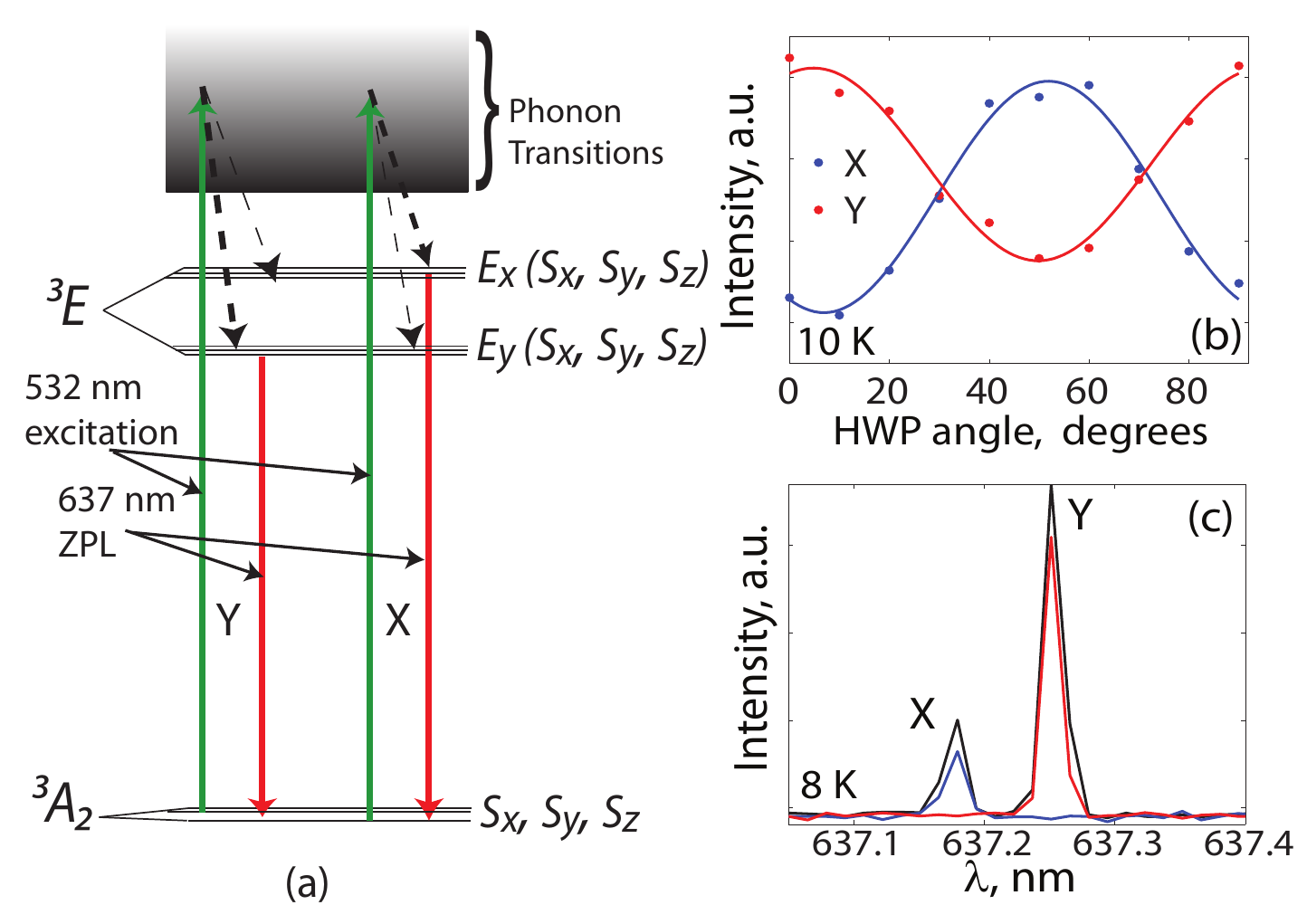}}\caption{(a) Electronic energy level diagram for the NV center in the presence of a strain field. (b ) PL intensity of the $E_x$ and $E_y$ peaks as a function of half waveplate (HWP) angle for NV3. (c) ZPL spectra of a single NV excited with Y-polarization. \emph{black}:  Collecting both polarizations. \emph{red}: Collecting Y polarization. \emph{blue}: Collecting X polarization.} \label{fig:energy}
\end{figure}

The NV center in diamond has trigonal symmetry (point group C3v) and consists of a single substitutional nitrogen with a nearest neighbor carbon vacancy. The NV axis ($z$ axis) can point along any of the four $\langle 111\rangle$ crystallographic axes. The ground state has $^3\!A_2$ symmetry (spin-triplet/orbital-singlet) and is split into an ${S_x,\, S_y}$ doublet $2.87 \, \mathrm{GHz}$ above an $S_z$ singlet.  These are connected by optical transitions to the spin-triplet/orbital-doublet $^3\!E$ excited state.
The excited-state structure involves spin-orbit and spin-spin interactions as well as a linear strain splitting of the two orbital states~\cite{ref:Manson2006nvc}.  The diagram in Fig.~\ref{fig:energy}a applies to the case where strain dominates over these other interactions such that the two orbital branches $E_x$ and $E_y$ are well separated in energy. Here, $x$ and $y$ are mutually orthogonal axes in a plane perpendicular to the NV axis with an angle determined by the strain tensor~\cite{ref:hughes1967uss}.  The ZPL optical transitions to $E_x$ and $E_y$ have orthogonal, linear polarization selection rules~\cite{ref:Davies1976os1} and are labeled X and Y. Excitation along the NV axis with X-polarized light should therefore produce X-polarized photoluminescence (PL), unless either population relaxes between the states, or else the selection rules are degraded, as occurs for excitation or collection through the phonon sidebands.

In the first experiment we measured the ZPL emission polarization as a function of temperature.  Five centers were studied in a $\langle111\rangle$-oriented type IIa natural diamond sample chosen for its low NV density.  The studied centers had their NV ($z$) axes oriented parallel to the excitation and collection path, as determined from their polarization anisotropy~\cite{ref:Epstein2005ais,ref:Alegre2007pse}. For NV1-2, no external stress was applied, but random strain fields were present due to local defects in the crystal.  For NV3-5, external stress was applied by mounting the sample between copper plates which contracted during the cryogenic cooldown process.  All experiments were performed in a coldfinger vacuum cryostat.

The polarization visibility $V = (I_\textrm{X}-I_\text{Y})/(I_\textrm{X}+I_\textrm{Y})$, in which $I_{\textrm{X(Y)}}$ is the intensity of the X(Y) polarized emission from the $^3\!E$ state, was measured for both X and Y polarized 532 nm excitation.  Alignment of linearly polarized excitation light to the principal axes of the NV center was performed by using a `pump' half waveplate directly before the final imaging objective.  To determine the proper waveplate angle, the X/Y transition intensity was measured as a function of waveplate angle as shown in Fig.~\ref{fig:energy}b.  The intensity extrema occur in this curve when the excitation field is either X or Y-polarized.
In a system with ideal polarization selection rules without excited-state relaxation, $V$ at these extrema should be equal to one, but we measure $|V|<0.5$ in all cases. A polarizer inserted before the spectrometer confirmed that the emission lines are in fact well polarized as shown in Fig.~\ref{fig:energy}c.  This, combined with the modest visibility temperature dependence at $10\,\mathrm{K}$ (see below), indicates that the main factor limiting the visibility in Fig.~\ref{fig:energy}b is imperfect \emph{excitation} selection rules when exciting non-resonantly through the high energy phonon sidebands.

\begin{figure}
\centering
{\includegraphics[width = 3.3in,
keepaspectratio]{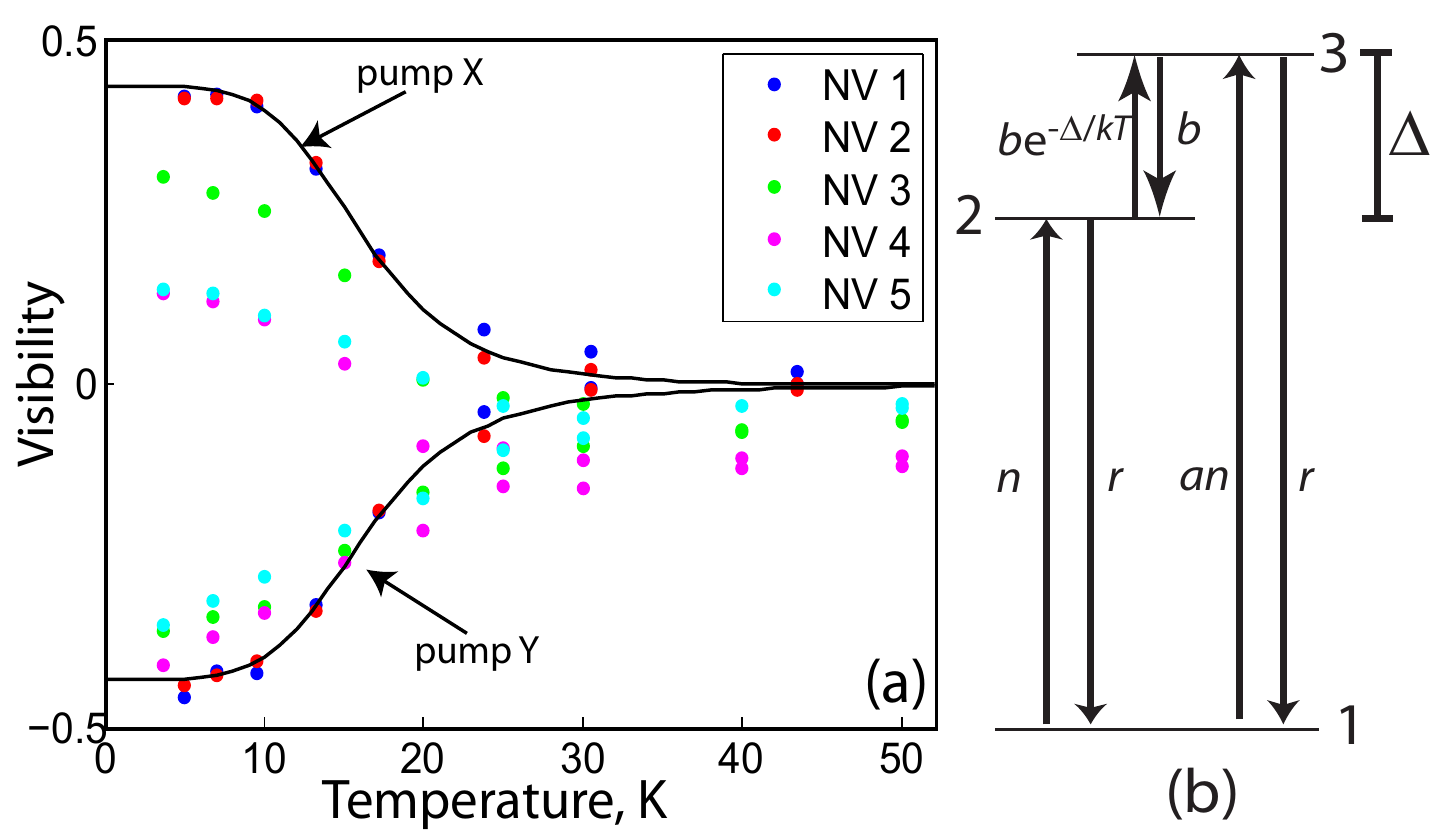}}\caption{(a) Temperature dependence of the polarization visibility $V = (I_\textrm{X}-I_\text{Y})/(I_\textrm{X}+I_\textrm{Y})$ for 5 NVs.  The solid black curve corresponds to a fit of NV1-2 data to the three-level model depicted in (b). For the spectrally \emph{resolved} NV3-5, the determination of the strain axis is discussed in the text. For NV1-2, an `analyzer' half waveplate followed by a fixed polarizer was inserted before the spectrometer to measure both components of the spectrally \emph{unresolved} X/Y ZPL peaks. The stain axis was determined by measuring the X/Y intensities obtained from rotating the `analyzer' waveplate for specific angles of the `pump' waveplate. Strain splittings for NV1-5 were 8, 9, 44, 54 and 81~GHz respectively. (b) Three level model used to fit the temperature dependent polarization relaxation for Y-polarized excitation.  For X-polarized excitation, $a$ and $an$ are reversed.} \label{fig:relax}
\end{figure}

The polarization visibility, plotted in Fig.~\ref{fig:relax}a, shows a strong temperature dependence from $10-40\,\mathrm{K}$ indicating that in this range the polarization decay rate is comparable to the radiative decay rate. A fit of the low strain NV1-2 polarization data to a simple three level model is shown in Fig.~\ref{fig:relax}a.  In this model, depicted in Fig.~\ref{fig:relax}b, the two excited states are split by energy $\Delta$ and the radiative decay rate is $r=(12.5 \, \mathrm{ns})^{-1}$~\cite{ref:batalov2008tcp, ref:Collins1983ldt}.  The model includes the excitation rate $n$ to the selected excited state plus a rate $a\times n$ to the other excited state.  The relaxation rate $b$ from states 3$\rightarrow$2 is taken to have a temperature dependence $b/r = c_1 T^N$.  The temperature dependence for each NV is fitted using parameters $c_1$, $N$ and $a$, and good agreement with experiment is obtained yielding $c_1 = (5.1\pm4.2) \times 10^{-7} \, \mathrm{K}^{-N}$, $N = 5.0 \pm 0.3$ and $a = 0.40\pm0.01$~\cite{errornoteFuPRLpol}.

In the second experiment we measured the linewidth of the ZPL through photoluminescence excitation (PLE) spectroscopy. In this measurement, photoluminescence into the phonon sidebands was detected as a tunable external-cavity diode laser operating at $637 \, \mathrm{nm}$ was scanned across the ZPL resonances.  Before each scan, a $532 \, \mathrm{nm}$ excitation pulse was applied to reverse photo-ionization which eventually occurs with $637 \, \mathrm{nm}$ excitation alone.  These measurements were performed on a commercially-obtained $\langle100\rangle$-oriented synthetic diamond sample (Element 6, Electronic Grade CVD) chosen for its high purity and relatively small spectral diffusion.  Two NV centers were studied, labeled NV 6 and 7.

\begin{figure}
\centering
\includegraphics[width=3.1in]{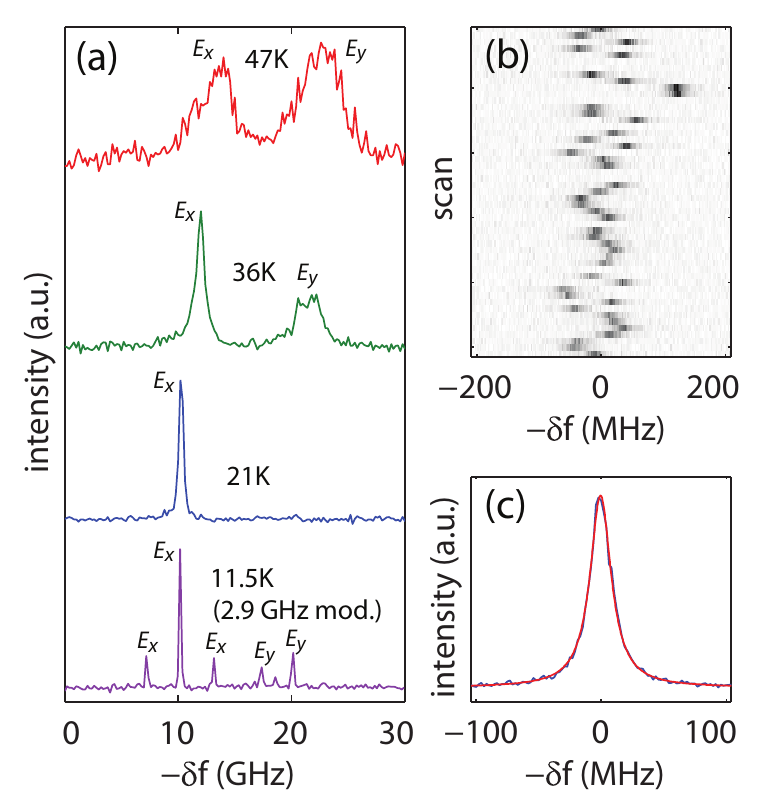}
\caption{(a) PLE spectra from NV7 at several temperatures.  Transitions are labeled according to the excited orbital state. (b) Individual PLE scans of NV7 illustrating spectral diffusion.  Intensity scale: 0 (white) to $8600\,\mathrm{s}^{-1}$ (black).  (c) Sum of scans (blue) using the procedure described in the text to remove spectral diffusion, and a Lorentzian fit (red).} \label{fig:ple_details}
\end{figure}

At the lowest temperatures, with a single excitation frequency, typically only a single PLE line is seen corresponding to the $m_s = 0$, $E_x$ transition, while all other transitions are hidden due to optical pumping~\cite{ref:Santori2006cpt,ref:Tamarat2008sfs,ref:Batalov2009lts}.  Transitions to the lower orbital branch ($E_y$) can be revealed by applying $2.9 \, \mathrm{GHz}$ modulation to the excitation laser, as shown in the bottom plot of Fig.~\ref{fig:ple_details}a.  The peaks at $\sim 10 \pm 2.9 \, \mathrm{GHz}$ correspond to the $m_s = 0$, $E_x$ transition excited by the laser and modulation sidebands, while the peaks at $\sim 17$ and $20 \, \mathrm{GHz}$ represent two situations in which the laser and modulation sidebands can simultaneously drive $m_s = 0$ and $m_s = \pm 1$, $E_y$ transitions.

At the lowest temperatures, linewidth estimation is complicated by spectral jumps and blinking of the PLE resonance, illustrated in Fig.~\ref{fig:ple_details}b. To remove this effect from the calculated linewidth, individual scans were shifted according to their first moment before summing.  A combined spectrum obtained in this way is shown in Fig.~\ref{fig:ple_details}c.  The raw linewidth was then obtained from a Lorentzian fit, also shown.  To avoid photobleaching and power broadening we used weak excitation powers: for example for NV 7 at low temperature the excitation power was $17 \, \mathrm{nW}$ (focused to a $1 \, \mu\mathrm{m}$ spot), while the saturation power was estimated to be $59 \, \mathrm{nW}$.  Power broadening corrections were applied to the raw linewidths, but these never exceeded 16\%.

\begin{figure}
\centering
\includegraphics[width=3.0in]{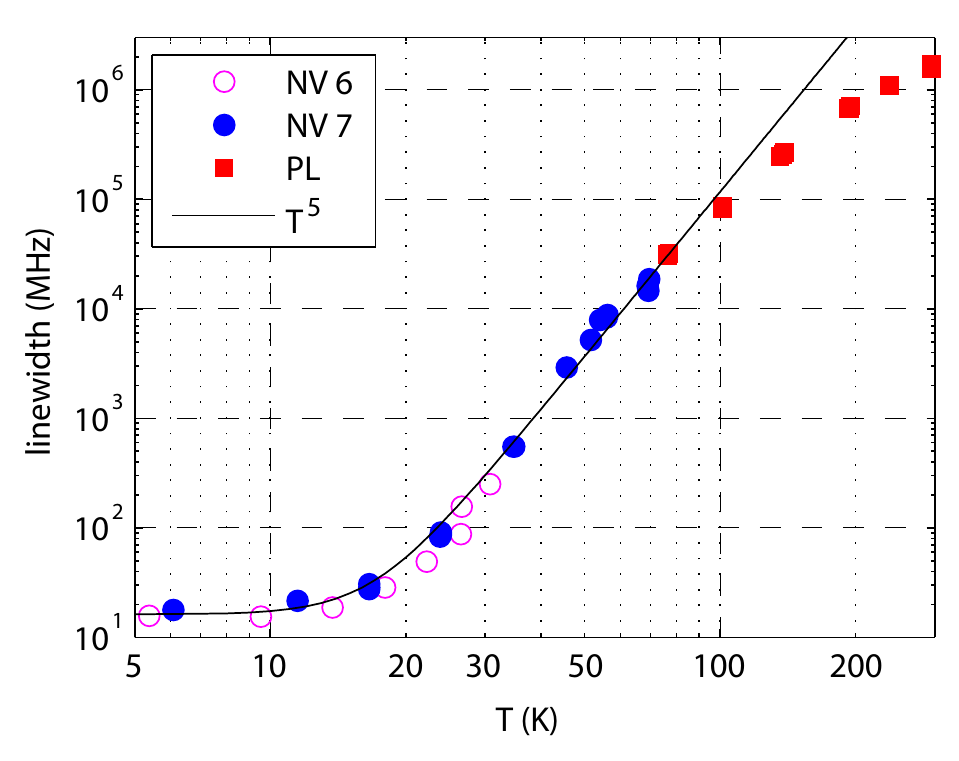}
\caption{Log-log plot showing the corrected PLE linewidths vs. temperature from NV6 and NV7, as well as PL linewidths from NV3-5 (PL).  The solid line is the $T^5$ fit described in the text.  The strain splitting for NV7 was 7.2~GHz and was undetermined for NV6.} \label{fig:ple_tdep}
\end{figure}

Linewidth results for two NV centers in the CVD sample are summarized in Fig.~\ref{fig:ple_tdep}.  For $T<10\,\mathrm{K}$ the linewidth is fairly constant and not much larger than the spontaneous emission lifetime limit of $13 \, \mathrm{MHz}$.  However, for $T>10\,\mathrm{K}$ the linewidth increases rapidly, following approximately a $T^5$ dependence.  The solid curve shows a fit using $\gamma(T) = \gamma_0 + c_2 r T^5$ up to 100~K with $\gamma_0 = 2\pi \times 16.2 \, \mathrm{MHz}$, $c_2 = (9.2\pm0.5) \times 10^{-7} \, \mathrm{K}^{-5}$, and $r = (12.5\,\mathrm{ns})^{-1}$ as above~\cite{errornoteFuPRLwidth}.  Fig.~\ref{fig:ple_tdep} also includes spectrometer-based PL linewidth measurements obtained from single NV centers, extending the experimental results to $300 \, \mathrm{K}$.
 Comparing the polarization relaxation and line broadening results, we find that the coefficients $c_1$ and $c_2$ differ only by a factor of two.  In these measurements this difference is not statistically significant, however we note that one possible cause for a difference could be the different strain environments in the two samples.

To explain this observed $T^5$ dependence, we propose a two-phonon Raman scattering process mediated by a linear electron-phonon interaction. The Hamiltonian is given by $H = H_e + H_v + H'$ in which $H_e$ is the NV electronic Hamiltonian and $H_v$ is the vibronic Hamiltonian.  $H'$ is the electron-phonon interaction and in the weak coupling limit can be written in the symmetry adapted form (excluding A$_2$ vibrations for brevity) ~\cite{ref:Ham1972svm}

\begin{eqnarray}
H' & = & \sum_{k} V^{A_1}_{k}\mathcal{U}_{a_1} (a_{ka_1} + a^\dag_{ka_1}) \nonumber \\  & &  + \sum_k  V^E_k(\mathcal{U}_{x}(a_{kx} + a^\dag_{kx}) + \mathcal{U}_{y} (a_{ky} + a^\dag_{ky}))
\label{eq:H'}
\end{eqnarray}
in which the first term relates to the interaction with A$_1$ vibrations  and the second with $E$ vibrations. $V_k^\Gamma$ gives the strength of the coupling to distortions of $\Gamma$ symmetry for phonon momentum $k$. Operator $a$ ($a^\dag$) is the phonon annihilation (creation) operator. $\mathcal{U}_{\gamma}$ are electronic operators transforming as row $\gamma$ of $\Gamma$ symmetry.  For strain applied along the NV's reflection symmetry axis, $\mathcal{U}_{\gamma}$ in the $\{E_x, E_y\}$ basis have the values ~\cite{ref:Ham1972svm}
$$
\begin{array}{ccccccccc}
\mathcal{U}_{a_1} & = & \left(\begin{array}{cc} 1 & 0 \\ 0 & 1 \end{array} \right), &
\mathcal{U}_{x} & = & \left(\begin{array}{cc} 1 & 0 \\ 0 & -1 \end{array} \right), & \mathcal{U}_{y} & = & -\left(\begin{array}{cc} 0 & 1 \\ 1 & 0 \end{array} \right).
\end{array}
$$

Considering the wavefunctions to first order in the electron-vibration interaction, for the case of $A_1$ vibrations there is no mixing between the $E_x$ and $E_y$  electronic states and as a consequence the polarization selection rules for the $A_1$ phonon sidebands are the same as for the ZPL.  For vibrations of $E$ symmetry, it can be seen from $\mathcal{U}_y$ that an $E_x\leftrightarrow E_y$ admixture does occur and the emitted polarization can be reversed.  Thus, the finite polarization visibility observed in Fig.~\ref{fig:relax}c at very low temperatures indicates that there is a significant electron-vibration interaction associated with vibrations of $E$ symmetry.  This is the first evidence of a DJT effect.

At  higher temperatures the terms in the electron-vibration interaction associated with $E$ vibrations can also induce transitions between the E$_x$ and E$_y$ states through a two-phonon Raman process. By using a symmetry adapted approach the rate for a transition $E_y\rightarrow E_x$ with an increase in a quantum of  $y$-vibration and decrease of a $x$-vibration is given by
\begin{eqnarray}
W_{xy} & = &2\pi\int_0^\infty d\omega\, n_\omega(n_\omega+1)\rho_{E_x}(\omega)\rho_{E_y}(\omega) \nonumber \\ & & \times V_\omega^4 \left|\left(\frac{\mathcal{U}_{xxx} \mathcal{U}_{xyy}}{-\omega}+\frac{\mathcal{U}_{xyy} \mathcal{U}_{yxy}}{+\omega}\right)\right|^2
\label{eq:rate}
\end{eqnarray}
in the limit $|E_x-E_y|\ll kT$. In Eq.~\ref{eq:rate}, $\mathcal{U}_{i l j} = \langle E_i|U_l|E_j \rangle$ and the phonon occupation number is given by $n_\omega = (\exp[\hbar\omega/k T]-1)^{-1}$ for phonon energy $\omega$. The phonon density of states of the $x$- and $y$-vibrations, given by $\rho_{E_x}(\omega)$ = $\rho_{E_y}(\omega)$, are both proportional to $\omega^2$ in the Debye approximation. The linear electron-phonon coupling $V_{\omega}$ varies as $\sqrt\omega$ in the absence of resonant phonon modes~\cite{ref:Maradudin1966pdd}.  Thus, the integrand in $W_{xy}$ varies as $\omega^4$ which leads to a $T^5$ dependence~\cite{ref:Walker1968tsl}. This is very different from the Raman rate associated with the coupling of vibrations to non-degenerate states which varies as $T^7$~\cite{ref:Orbach1962slr}. The experimentally observed $T^5$ dependence in the NV polarization experiment is the second evidence for  coupling to a degenerate vibrational mode, or a DJT effect, in the low strain NV system. When the strain splitting is larger, as with NV3-5, the strain further degrades the selection rules and in addition one-phonon relaxation results in an asymmetry in the visibility curves for the two pump polarizations.  This is qualitatively what is observed in Fig.~\ref{fig:relax}c.

The Raman relaxation shortens the electronic lifetime in either $^3\!E$ state level resulting in a dephasing of the optical transition and an increased optical linewidth. The observed $T^5$ broadening of the ZPL and the similar magnitudes of $c_1$ and $c_2$ strongly suggest that the DJT effect also accounts for the optical dephasing at low T.  If the linewidth is solely determined by the $^3\!E$ population relaxation we would expect $c_1 = c_2 = 2W_{xy}/rT^5.$

The established presence of a DJT effect in the negatively charged NV system is significant as the effect is known to substantially diminish interactions such as spin-orbit, orbital Zeeman, and the response to external perturbations such as stress and electric field ~\cite{ref:Ham1965djt,ref:Ham1972svm}.  It can, therefore, be understood why the effects of many of these interactions are found to be small in the NV system. The DJT effect may also play a role in the fast population transfer between the excited states observed at room temperature. This fast transfer effectively averages the two orbital states resulting in a significant simplification of the excited state spin resonance structure observed at room temperature~\cite{ref:Rogers2009taw, ref:Fuchs2008ess, ref:Neumann2009ess}.  Finally, by improving our basic understanding of the primary decoherence mechanism affecting the optical transitions, these results should allow much more accurate performance estimates for quantum information processing applications including schemes designed to mitigate the effects of excited-state dephasing~\cite{ref:Santori2009irp}.

This material is based upon work supported by the Defense Advanced Research Projects Agency under Award No. HR0011-09-1-0006 and The Regents of the University of California, and the Australian Research Council.

\bibliographystyle{apsrev}

\begin{thebibliography}{32}
\expandafter\ifx\csname natexlab\endcsname\relax\def\natexlab#1{#1}\fi
\expandafter\ifx\csname bibnamefont\endcsname\relax
  \def\bibnamefont#1{#1}\fi
\expandafter\ifx\csname bibfnamefont\endcsname\relax
  \def\bibfnamefont#1{#1}\fi
\expandafter\ifx\csname citenamefont\endcsname\relax
  \def\citenamefont#1{#1}\fi
\expandafter\ifx\csname url\endcsname\relax
  \def\url#1{\texttt{#1}}\fi
\expandafter\ifx\csname urlprefix\endcsname\relax\def\urlprefix{URL }\fi
\providecommand{\bibinfo}[2]{#2}
\providecommand{\eprint}[2][]{\url{#2}}

\bibitem[{\citenamefont{{G. Balasubramanian \emph{et
  al.}}}(2009)}]{ref:Balasubramanian2009usc}
\bibinfo{author}{\bibnamefont{{G. Balasubramanian \emph{et al.}}}},
  \bibinfo{journal}{Nature Materials} \textbf{\bibinfo{volume}{8}},
  \bibinfo{pages}{383} (\bibinfo{year}{2009}).

\bibitem[{\citenamefont{{T. Gaebel \emph{et. al.}}}(2006)}]{ref:Gaebel2006rtc}
\bibinfo{author}{\bibnamefont{{T. Gaebel \emph{et. al.}}}},
  \bibinfo{journal}{Nature Phys.} \textbf{\bibinfo{volume}{2}},
  \bibinfo{pages}{408} (\bibinfo{year}{2006}).

\bibitem[{\citenamefont{{M. V. Gurudev Dutt \emph{et
  al.}}}(2007)}]{ref:GurudevDutt2007qua}
\bibinfo{author}{\bibnamefont{{M. V. Gurudev Dutt \emph{et al.}}}},
  \bibinfo{journal}{Science} \textbf{\bibinfo{volume}{316}},
  \bibinfo{pages}{1312} (\bibinfo{year}{2007}).

\bibitem[{\citenamefont{{P. Neumann \emph{et al.}}}(2008)}]{ref:Neuman2008mea}
\bibinfo{author}{\bibnamefont{{P. Neumann \emph{et al.}}}},
  \bibinfo{journal}{Science} \textbf{\bibinfo{volume}{320}},
  \bibinfo{pages}{1326} (\bibinfo{year}{2008}).

\bibitem[{\citenamefont{{R. Hanson \emph{et al.}}}(2008)}]{ref:Hanson2008cds}
\bibinfo{author}{\bibnamefont{{R. Hanson \emph{et al.}}}},
  \bibinfo{journal}{Science} \textbf{\bibinfo{volume}{320}},
  \bibinfo{pages}{352} (\bibinfo{year}{2008}).

\bibitem[{\citenamefont{{J. M. Taylor \emph{et
  al.}}}(2008)}]{ref:Taylor2008hig}
\bibinfo{author}{\bibnamefont{{J. M. Taylor \emph{et al.}}}},
  \bibinfo{journal}{Nature Phys.} \textbf{\bibinfo{volume}{4}},
  \bibinfo{pages}{810} (\bibinfo{year}{2008}).

\bibitem[{\citenamefont{{C. Santori \emph{et al.}}}(2006)}]{ref:Santori2006cpt}
\bibinfo{author}{\bibnamefont{{C. Santori \emph{et al.}}}},
  \bibinfo{journal}{Phys. Rev. Lett.} \textbf{\bibinfo{volume}{97}},
  \bibinfo{pages}{247401} (\bibinfo{year}{2006}).

\bibitem[{\citenamefont{{P. Tamarat \emph{et al.}}}(2008)}]{ref:Tamarat2008sfs}
\bibinfo{author}{\bibnamefont{{P. Tamarat \emph{et al.}}}},
  \bibinfo{journal}{New J. Phys.} \textbf{\bibinfo{volume}{10}},
  \bibinfo{pages}{045004} (\bibinfo{year}{2008}).

\bibitem[{\citenamefont{{S. C. Benjamin \emph{et
  al.}}}(2006)}]{ref:Benjamin2006bgs}
\bibinfo{author}{\bibnamefont{{S. C. Benjamin \emph{et al.}}}},
  \bibinfo{journal}{New J. of Phys.} \textbf{\bibinfo{volume}{8}},
  \bibinfo{pages}{141} (\bibinfo{year}{2006}).

\bibitem[{\citenamefont{{S. M. Clark \emph{et al.}}}(2007)}]{ref:Clark2007qcb}
\bibinfo{author}{\bibnamefont{{S. M. Clark \emph{et al.}}}},
  \bibinfo{journal}{Phys. Rev. Lett.} \textbf{\bibinfo{volume}{99}},
  \bibinfo{pages}{040501} (\bibinfo{year}{2007}).

\bibitem[{\citenamefont{{T. D. Ladd \emph{et al.}}}(2006)}]{ref:Ladd2006hqr}
\bibinfo{author}{\bibnamefont{{T. D. Ladd \emph{et al.}}}},
  \bibinfo{journal}{New Journal of Physics} \textbf{\bibinfo{volume}{8}},
  \bibinfo{pages}{184} (\bibinfo{year}{2006}).

\bibitem[{\citenamefont{{L. Childress \emph{et
  al.}}}(2005)}]{ref:Childress2005ftq}
\bibinfo{author}{\bibnamefont{{L. Childress \emph{et al.}}}},
  \bibinfo{journal}{Phys. Rev.~A} \textbf{\bibinfo{volume}{72}},
  \bibinfo{pages}{052330} (\bibinfo{year}{2005}).

\bibitem[{\citenamefont{Davies}(1974)}]{ref:Davies1974vsd}
\bibinfo{author}{\bibfnamefont{G.}~\bibnamefont{Davies}}, \bibinfo{journal}{J.
  Phys.C: Solid State Phys.} \textbf{\bibinfo{volume}{7}},
  \bibinfo{pages}{3797} (\bibinfo{year}{1974}).

\bibitem[{\citenamefont{Maradudin}(1966)}]{ref:Maradudin1966pdd}
\bibinfo{author}{\bibfnamefont{A.~A.} \bibnamefont{Maradudin}},
  \emph{\bibinfo{title}{Solid State Physics}} (\bibinfo{publisher}{Academic
  Press}, \bibinfo{address}{New York}, \bibinfo{year}{1966}),
  vol.~\bibinfo{volume}{18}, pp. \bibinfo{pages}{273--420}.

\bibitem[{\citenamefont{Manson et~al.}(2006)\citenamefont{Manson, Harrison, and
  Sellars}}]{ref:Manson2006nvc}
\bibinfo{author}{\bibfnamefont{N.~B.} \bibnamefont{Manson}},
  \bibinfo{author}{\bibfnamefont{J.~P.} \bibnamefont{Harrison}},
  \bibnamefont{and} \bibinfo{author}{\bibfnamefont{M.~J.}
  \bibnamefont{Sellars}}, \bibinfo{journal}{Phys. Rev.~B}
  \textbf{\bibinfo{volume}{74}}, \bibinfo{pages}{104303}
  (\bibinfo{year}{2006}).

\bibitem[{\citenamefont{Hughes and Runciman}(1967)}]{ref:hughes1967uss}
\bibinfo{author}{\bibfnamefont{A.}~\bibnamefont{Hughes}} \bibnamefont{and}
  \bibinfo{author}{\bibfnamefont{W.}~\bibnamefont{Runciman}},
  \bibinfo{journal}{Proc. Phys. Soc. London} \textbf{\bibinfo{volume}{90}},
  \bibinfo{pages}{827} (\bibinfo{year}{1967}).

\bibitem[{\citenamefont{Davies and Hamer}(1976)}]{ref:Davies1976os1}
\bibinfo{author}{\bibfnamefont{G.}~\bibnamefont{Davies}} \bibnamefont{and}
  \bibinfo{author}{\bibfnamefont{M.~F.} \bibnamefont{Hamer}},
  \bibinfo{journal}{Proc. R. Soc. A} \textbf{\bibinfo{volume}{348}},
  \bibinfo{pages}{285} (\bibinfo{year}{1976}).

\bibitem[{\citenamefont{{R. J. Epstein \emph{et
  al.}}}(2005)}]{ref:Epstein2005ais}
\bibinfo{author}{\bibnamefont{{R. J. Epstein \emph{et al.}}}},
  \bibinfo{journal}{Nature Phys.} \textbf{\bibinfo{volume}{1}},
  \bibinfo{pages}{94} (\bibinfo{year}{2005}).

\bibitem[{\citenamefont{{T.P.{Mayer Alegre} \emph{et
  al.}}}(2007)}]{ref:Alegre2007pse}
\bibinfo{author}{\bibnamefont{{T.P.{Mayer Alegre} \emph{et al.}}}},
  \bibinfo{journal}{Phys. Rev.~B} \textbf{\bibinfo{volume}{76}},
  \bibinfo{pages}{165205} (\bibinfo{year}{2007}).

\bibitem[{\citenamefont{{A. Batalov \emph{et al.}}}(2008)}]{ref:batalov2008tcp}
\bibinfo{author}{\bibnamefont{{A. Batalov \emph{et al.}}}},
  \bibinfo{journal}{Phys. Rev. Lett.} \textbf{\bibinfo{volume}{100}},
  \bibinfo{pages}{077401} (\bibinfo{year}{2008}).

\bibitem[{\citenamefont{Collins et~al.}(1983)\citenamefont{Collins, Thomaz, and
  Jorge}}]{ref:Collins1983ldt}
\bibinfo{author}{\bibfnamefont{A.~T.} \bibnamefont{Collins}},
  \bibinfo{author}{\bibfnamefont{M.~F.} \bibnamefont{Thomaz}},
  \bibnamefont{and} \bibinfo{author}{\bibfnamefont{M.~I.~B.}
  \bibnamefont{Jorge}}, \bibinfo{journal}{J. Phys. C}
  \textbf{\bibinfo{volume}{16}}, \bibinfo{pages}{2177} (\bibinfo{year}{1983}).

\bibitem[{err({\natexlab{a}})}]{errornoteFuPRLpol}
\bibinfo{note}{Fitting error includes only Poisson noise in the measured
  intensities. Taking into account possible systematic errors in the
  temperature measurement, the exponent uncertainty increases to $N = 5.0 +1.0/
  -0.3$.}

\bibitem[{\citenamefont{{A. Batalov \emph{et al.}}}(2009)}]{ref:Batalov2009lts}
\bibinfo{author}{\bibnamefont{{A. Batalov \emph{et al.}}}},
  \bibinfo{journal}{Phys. Rev. Lett.} \textbf{\bibinfo{volume}{102}},
  \bibinfo{pages}{195506} (\bibinfo{year}{2009}).

\bibitem[{err({\natexlab{b}})}]{errornoteFuPRLwidth}
\bibinfo{note}{Error is derived from observed fluctuations in the data.}

\bibitem[{\citenamefont{Ham}(1972)}]{ref:Ham1972svm}
\bibinfo{author}{\bibfnamefont{F.~S.} \bibnamefont{Ham}},
  \bibinfo{journal}{Phys. Rev. Lett.} \textbf{\bibinfo{volume}{28}},
  \bibinfo{pages}{1048} (\bibinfo{year}{1972}).

\bibitem[{\citenamefont{Walker}(1968)}]{ref:Walker1968tsl}
\bibinfo{author}{\bibfnamefont{M.~B.} \bibnamefont{Walker}},
  \bibinfo{journal}{Can. J. of Phys.} \textbf{\bibinfo{volume}{46}},
  \bibinfo{pages}{1347} (\bibinfo{year}{1968}).

\bibitem[{\citenamefont{Orbach and Blume}(1962)}]{ref:Orbach1962slr}
\bibinfo{author}{\bibfnamefont{R.}~\bibnamefont{Orbach}} \bibnamefont{and}
  \bibinfo{author}{\bibfnamefont{M.}~\bibnamefont{Blume}},
  \bibinfo{journal}{Phys. Rev. Lett.} \textbf{\bibinfo{volume}{8}},
  \bibinfo{pages}{478} (\bibinfo{year}{1962}).

\bibitem[{\citenamefont{Ham}(1965)}]{ref:Ham1965djt}
\bibinfo{author}{\bibfnamefont{F.~S.} \bibnamefont{Ham}},
  \bibinfo{journal}{Phys. Rev.} \textbf{\bibinfo{volume}{138}},
  \bibinfo{pages}{A1727} (\bibinfo{year}{1965}).

\bibitem[{\citenamefont{{L. J. Rogers \emph{et
  al.}}}(2009)}]{ref:Rogers2009taw}
\bibinfo{author}{\bibnamefont{{L. J. Rogers \emph{et al.}}}},
  \bibinfo{journal}{New J. Phys.} \textbf{\bibinfo{volume}{11}},
  \bibinfo{pages}{063007} (\bibinfo{year}{2009}).

\bibitem[{\citenamefont{{G. D. Fuchs \emph{et al.}}}(2008)}]{ref:Fuchs2008ess}
\bibinfo{author}{\bibnamefont{{G. D. Fuchs \emph{et al.}}}},
  \bibinfo{journal}{Phys. Rev. Lett.} \textbf{\bibinfo{volume}{101}},
  \bibinfo{pages}{117601} (\bibinfo{year}{2008}).

\bibitem[{\citenamefont{{P. Neumann \emph{et al.}}}(2009)}]{ref:Neumann2009ess}
\bibinfo{author}{\bibnamefont{{P. Neumann \emph{et al.}}}},
  \bibinfo{journal}{New J. Phys.} \textbf{\bibinfo{volume}{11}},
  \bibinfo{pages}{013017} (\bibinfo{year}{2009}).

\bibitem[{\citenamefont{{ C.Santori \emph{et al.} }}()}]{ref:Santori2009irp}
\bibinfo{author}{\bibnamefont{{ C.Santori \emph{et al.} }}},
  \bibinfo{journal}{New J. Phys.} \textbf{\bibinfo{volume}{11}},
  \bibinfo{pages}{123009} (\bibinfo{year}{2009}).

\end{thebibliography}

\end{document}